\documentclass[aps,axodraw,twocolumn,prl,nofootinbib,superscriptaddress]{revtex4}
\usepackage{fancyhdr}
\usepackage{fancybox}
\usepackage{epstopdf}

\usepackage{graphicx}
\usepackage{color}
\usepackage{soul}
\usepackage{latexsym}

\def\metab{m_{\eta_b}}
\def\cta{\cos\theta_A}
\def\mupsilon{M_\Upsilon}
\def\tauptaum{\tau^+\tau^-}
\def\brups{\br(\Upsilon\to \gam\ai)}

\def\ma{m_a}
\def\mh{m_h}
\def\hi{h_1}

\def\ai{a_1}
\def\mhi{m_{h_1}}

\def\mai{m_{a_1}}
\def\mueff{\mu_{\rm eff}}
\def\mtau{m_\tau}

\def\lam{\lambda}
\def\kap{\kappa}
\def\alam{A_\lambda}
\def\akap{A_\kappa}

\def\lam{\lambda}

\def\mh{m_h}

\def\h{h}
\def\mh{m_{h}}

\def\what{\widehat}

\def\hbar{\overline h}
\def\lam{\lambda}

\def\mz{m_Z}

\def\hi{h_i^0}
\def\mhi{m_{\hi}}

\def\h{h}
\def\mh{m_{\h}}

\def\lam{\lambda}

\def\wtil{\widetilde}
\def\what{\widehat}
\def\tauptaum{\tau^+\tau^-}

\def\lsim{\mathrel{\raise.3ex\hbox{$<$\kern-.75em\lower1ex\hbox{$\sim$}}}}
\def\gsim{\mathrel{\raise.3ex\hbox{$>$\kern-.75em\lower1ex\hbox{$\sim$}}}}
\def\ifmath#1{\relax\ifmmode #1\else $#1$\fi}
\def\half{\ifmath{{\textstyle{1 \over 2}}}}

\def\third{\ifmath{{\textstyle{1 \over 3}}}}

\def\vev#1{\langle #1 \rangle}
\def\lam{\lambda}

\def\mhi{m_{h_1}}

\def\eg{{\it e.g.}}

\def\eg{{\it e.g.}}

\def\tanb{\tan\beta}

\def\mb{m_b}
\def\mz{m_Z}

\def\cnone{\wt\chi^0_1}

\def\mcnone{m_{\cnone}}

\def\wt{\widetilde}

\def\MPL #1 #2 #3 {{\sl Mod.~Phys.~Lett.}~{\bf#1} (#3) #2}
\def\NPB #1 #2 #3 {{\sl Nucl.~Phys.}~{\bf #1} (#3) #2}
\def\PLB #1 #2 #3 {{\sl Phys.~Lett.}~{\bf #1} (#3) #2}
\def\PR #1 #2 #3 {{\sl Phys.~Rep.}~{\bf#1} (#3) #2}
\def\PRD #1 #2 #3 {{\sl Phys.~Rev.}~{\bf #1} (#3) #2}
\def\PRL #1 #2 #3 {{\sl Phys.~Rev.~Lett.}~{\bf#1} (#3) #2}
\def\RMP #1 #2 #3 {{\sl Rev.~Mod.~Phys.}~{\bf#1} (#3) #2}
\def\ZPC #1 #2 #3 {{\sl Z.~Phys.}~{\bf #1} (#3) #2}
\def\IJMP #1 #2 #3 {{\sl Int.~J.~Mod.~Phys.}~{\bf#1} (#3) #2}
\def\NIM #1 #2 #3 {{\sl Nucl.~Inst.~and~Meth.}~{\bf#1} {#3} #2}

\def\lam{\lambda}
\def\br{{\rm Br}}
\def\tauptaum{\tau^+\tau^-}

\def\gam{\gamma}

\def\anti{\overline}
\def\epem{e^+e^-}
\def\mupmum{\mu^+\mu^-}

\def\mupmum{\mu^+\mu^-}

\def\ie{{\it i.e.}}
\def\eg{{\it e.g.}}
\def\eps{\epsilon}
\def\anti{\overline}

\def\ai{a_1}

\def\mai{m_{\ai}}

\def\mev{~{\rm MeV}}
\def\gev{~{\rm GeV}}

\def\mb{m_b}

\def\hi{\h_1}

\def\mhi{m_{\hi}}

\newcommand{\nc}{\newcommand}
\nc{\beq}{\begin{equation}}   \nc{\eeq}{\end{equation}}
\nc{\bea}{\begin{eqnarray}}   \nc{\eea}{\end{eqnarray}}
\nc{\baa}{\begin{array}}      \nc{\eaa}{\end{array}}
\nc{\bit}{\begin{itemize}}    \nc{\eit}{\end{itemize}}
\nc{\ben}{\begin{enumerate}}  \nc{\een}{\end{enumerate}}
\nc{\bce}{\begin{center}}     \nc{\ece}{\end{center}}
\def\beqa{\begin{eqnarray}}
\def\eeqa{\end{eqnarray}}
\def\bed{\begin{description}}
\def\eed{\end{description}}

\def\mhi{m_{h_1}}

\def\eg{{\it e.g.}}

\def\half{\frac{1}{2}\,}
\def\third{\frac{1}{3}\,}

\def\tanb{\tan\beta}

\def\simle{
    \mathrel{\rlap{\raise 0.511ex
        \hbox{$<$}}{\lower 0.511ex \hbox{$\sim$}}}}

\def\slashchar#1{\setbox0=\hbox{$#1$}           
   \dimen0=\wd0                                 
   \setbox1=\hbox{/} \dimen1=\wd1               
   \ifdim\dimen0>\dimen1                        
      \rlap{\hbox to \dimen0{\hfil/\hfil}}      
      #1                                        
   \else                                        
      \rlap{\hbox to \dimen1{\hfil$#1$\hfil}}   
      /                                         
   \fi}

\def\lam{\lambda}

\def\ls#1{\ifmath{_{\lower1.5pt\hbox{$\scriptstyle #1$}}}}
\def\lss#1{\ifmath{^{\,\lower2.5pt\hbox{$\scriptstyle #1$}}}}

\begin{document}

\preprint{UCD-2006-17}
\title{\boldmath 
Probing NMSSM Scenarios with Minimal Fine-Tuning by Searching for
Decays of the Upsilon to a Light CP-Odd Higgs Boson
}

\author{Radovan Derm\' \i\v sek}

\affiliation{School of Natural Sciences, Institute for Advanced Study, Princeton,
NJ 08540}

\author{John F. Gunion}

\affiliation{Department of Physics, University of California at Davis, Davis, CA 95616}

\author{Bob McElrath}

\affiliation{Department of Physics, University of California at Davis, Davis, CA 95616}

\date{\today}

\begin{abstract}
  
  Completely natural electroweak symmetry breaking is easily achieved
  in supersymmetric models if there is a SM-like Higgs boson, $h$,
  with $\mh\lsim 100\gev$.  In the minimal supersymmetric model, such
  an $h$ decays mainly to $b\anti b$ and is ruled out by LEP
  constraints. However, if the MSSM Higgs sector is expanded so that
  $h$ decays mainly to still lighter Higgs bosons, \eg\ $h\to aa$,
  with $\br(h\to aa)>0.7$, and if $\ma<2\mb$, then the LEP constraints
  are satisfied.  In this letter, we show that in the next-to-minimal
  supersymmetric model the above $h$ and $a$ properties (for the
  lightest CP-even and CP-odd Higgs bosons, respectively) imply a
  lower bound on $\br(\Upsilon\to \gam a)$ that dedicated runs at 
  present (and future) $B$ factories can explore.

\end{abstract}

\pacs{}
\keywords{}

\maketitle

Low energy supersymmetry remains one of the most attractive solutions
to the naturalness / hierarchy problem of the Standard Model (SM).
However, the minimal supersymmetric model (MSSM), containing exactly
two Higgs doublets, suffers from the ``$\mu$ problem'' and requires
rather special parameter choices in order that the light Higgs mass is
above LEP limits without electroweak symmetry breaking being
``fine-tuned'', \ie\ highly sensitive to supersymmetry-breaking
parameters chosen at the grand-unification scale. Both problems are
easily solved by adding Higgs (super) fields to the MSSM. For generic
SUSY parameters well-below the TeV scale, fine-tuning is
absent~\cite{Dermisek:2005ar} and a SM-like $h$ is predicted with
$\mh\lsim 100\gev$. Such an $h$ can avoid LEP limits on the tightly
constrained $\epem\to Z+b's$ channel if $\br(h\to b \anti b)$ is small
by virtue of large $\br(h\to aa)$, where $a$ is a new light (typically
CP-odd) Higgs boson, {\it and} $\ma<2\mb$ so that $a\to b\anti b$ is
forbidden~\cite{Dermisek:2005gg}. The perfect place to search for such
an $a$ is in Upsilon decays, $\Upsilon\to \gam a$.  The simplest MSSM
extension, the next-to-minimal supersymmetric model (NMSSM), naturally
predicts that the lightest $h$ and $a$, $\hi$ and $\ai$,
have all the right
features~\cite{Dermisek:2005ar,Dermisek:2005gg,newewsb,Dermisek:2006ya,Dermisek:2006wr}.
In this letter, we show that large $\br(\hi\to \ai\ai)$ implies, at
fixed $\mai$, a lower bound on $\brups$ (from now on, $\Upsilon$ is
the $1S$ resonance unless otherwise stated) that is typically within
reach of present and future $B$ factories.

In the NMSSM, a light $\ai$ with substantial $\br(\hi\to\ai\ai)$ is
a very natural possibility for $\mz$-scale soft parameters
developed by renormalization group running starting from $U(1)_R$
symmetric GUT-scale soft parameters~\cite{Dermisek:2006wr}. (See
also~\cite{Dobrescu:2000jt,Hiller:2004ii} for discussions of the light
$\ai$ scenario.)  The fine-tuning-preferred  
$\mhi\sim 100\gev$ (for $\tanb\gsim few$) gives perfect
consistency with precision electroweak data and the
reduced $\br(\hi\to b\anti b)\sim 0.09-0.15$ explains
the $\sim 2.3\sigma$ excess at LEP in the $Zb\anti b$
channel at $M_{b\anti b}\sim 100\gev$. The motivation for this
scenario is thus very strong.

Hadron collider probes of the NMSSM Higgs sector are problematical.
The $\hi\to \ai\ai\to 4\tau$ ($2\mtau<\mai<2\mb$) or $4~jets$
($\mai<2\mtau$) signal is a very difficult one at the Tevatron and
very possibly at the
LHC~\cite{Gunion:1996fb,Ellwanger:2003jt,Ellwanger:2004gz,Ellwanger:2005uu}.
Higgs discovery or, at the very least, certification of a marginal LHC
Higgs signal will require a linear $\epem$ collider (ILC).  Direct
production and detection of the $\ai$ may be impossible at both the
LHC and ILC because it is rather singlet in nature.  In this letter, we
show that by increasing sensitivity to $\brups$ 
by one to three orders of magnitude 
(the exact requirement depends on $\mai$ and $\tanb$), there is a good chance of 
detecting the $\ai$. This
constitutes a significant opportunity for current $B$ factories
and a major motivation for new super-$B$ factories.  Even with ILC $\hi\to\ai\ai$ data, measurement
of $\brups$ and $\ai$ decays would provide
extremely valuable complementary information.

As compared to the three independent parameters needed in the
MSSM context (often chosen as $\mu$, $\tan \beta$ and $M_A$), the
Higgs sector of the NMSSM is described by the six parameters
\vspace*{-.1in}
\beq \label{6param}
\lambda\ , \ \kappa\ , \ A_{\lambda} \ , \ A_{\kappa}, \ \tan \beta\ ,
\ \mu_\mathrm{eff}\ ,
\vspace*{-.1in}
  \eeq
where $\mueff=\lam\vev{S}\equiv \lam s$ is the effective $\mu$-term generated from the
$\lam\what S \what H_u \what H_d$ part of the superpotential,
$\lam\alam S H_u H_d$ is the associated soft-SUSY-breaking scalar potential
component, and $\kap$ and $\kap\akap$ appear in the 
$\third \kap \what S^3$ and $\third\kap\akap S^3$ terms in the superpotential and 
associated soft-supersymmetry-breaking potential.
  In addition, values must be input for the soft SUSY-breaking masses
  that contribute to the
  radiative corrections in the Higgs sector and to the Higgs decay
  widths.   Our computations for
  branching ratios and so forth employ
  NMHDECAY~\cite{Ellwanger:2004xm}. An important ingredient for the
  results of this paper is the non-singlet fraction of the $\ai$
  defined by $\cta$ in
\beq
\ai=\cta
A_{MSSM}+\sin\theta_A A_S\,,
\eeq
where $A_S$ is the CP-odd Higgs boson contained in the unmixed $S$
complex scalar field. The coupling of $\ai$ to $\tauptaum$ and $b\anti
b$ is then $\propto \tanb\cta$; $\cta$ itself has some $\tanb$
dependence with the net result that
$\tanb\cta$ increases modestly with increasing $\tanb$.

In~\cite{Dermisek:2005ar,newewsb,Dermisek:2006ya},
we scanned over the NMSSM parameter space
holding $\tanb$ and the gaugino masses $M_{1,2,3}(\mz)$ fixed, searching for choices
that minimized a numerical measure, $F$, of EWSB
fine-tuning, \ie\ of how precisely the GUT-scale
soft-SUSY-breaking parameters must be chosen to obtain the observed
value of $\mz$ after RG evolution.  For $F<15$,
fine-tuning is no worse than 7\%, and we regard this 
as equivalent to absence of
significant fine-tuning. For the sample
values of $\tanb=10$ and $M_{1,2,3}=100,200,300\gev$ ($F$ only depends
significantly on $M_3$), to
achieve the lowest $F$ values ($F\sim 5-6$), 
the $\hi$ must be fairly SM-like
and $\mhi\sim
100\gev$ is required; this is only consistent with LEP constraints
for scenarios in which $\br(\hi \to \ai \ai)$ is large
and $\mai < 2 m_b$.~\footnote{We should note that the
precise location of the minimum in $F$ shifts slightly as $\tanb$ is
varied. For example, at $\tanb=3$ ($\tanb=50$)  
the minimum is at roughly $92\gev$ ($102\gev$). However, for these cases 
the minimum value of $F$ is only very modestly
higher at $\mhi\sim 100\gev$, the LEP excess location.} Crucially, for
these scenarios one
finds  a lower bound on $|\cta|$, \eg\
$|\cta|\gsim 0.04$ at $\tanb=10$. As described
in~\cite{Dermisek:2006wr}, this is required in order that
$\br(\hi\to\ai\ai)>0.7$ when $\mai<2\mb$.~\footnote{Also, as one
  approaches the 
$U(1)_R$, $\akap,\alam\to 0$ symmetry limit, large $\br(\hi\to\ai\ai)$
is not possible.}

Aside from EWSB fine-tuning, there is a question of whether
fine-tuning is needed to achieve large $\br(\hi \to \ai \ai)$ 
and $\mai < 2 m_b$ when $F<15$.  This was discussed
in~\cite{Dermisek:2006wr}.  The level of such fine-tuning is 
determined mostly by whether $\alam$ and $\akap$ need to be
fine-tuned. (For given $s$ and $\tanb$,
$\br(\hi\to \ai\ai)$ and $\mai$ depend significantly 
only on $\lam$, $\kap$, $\alam$ and $\akap$;
all other SUSY parameters have only a tiny influence.)
Since specific soft-SUSY-breaking
scenarios can evade the issue of tuning $\akap$ and $\alam$
altogether, in this study we do not impose a limit on the
measures of $\alam,\akap$ fine-tuning discussed
in~\cite{Dermisek:2006wr}. However, it 
is worth noting that we find that $\alam,\akap$
fine-tuning can easily be avoided if $\mai\gsim 2\mtau$ 
and $\cta$ is small and negative, \eg\ near
$\cta\sim -0.05$ if $\tanb=10$. In some models, the simplest measures 
of $\alam,\akap$ fine-tuning are much larger
away from the preferred $\cta$ region and / or at
substantially lower $\mai$ values.  

\begin{figure}
\vspace*{.2in}
\includegraphics[width=2.6in,angle=90]{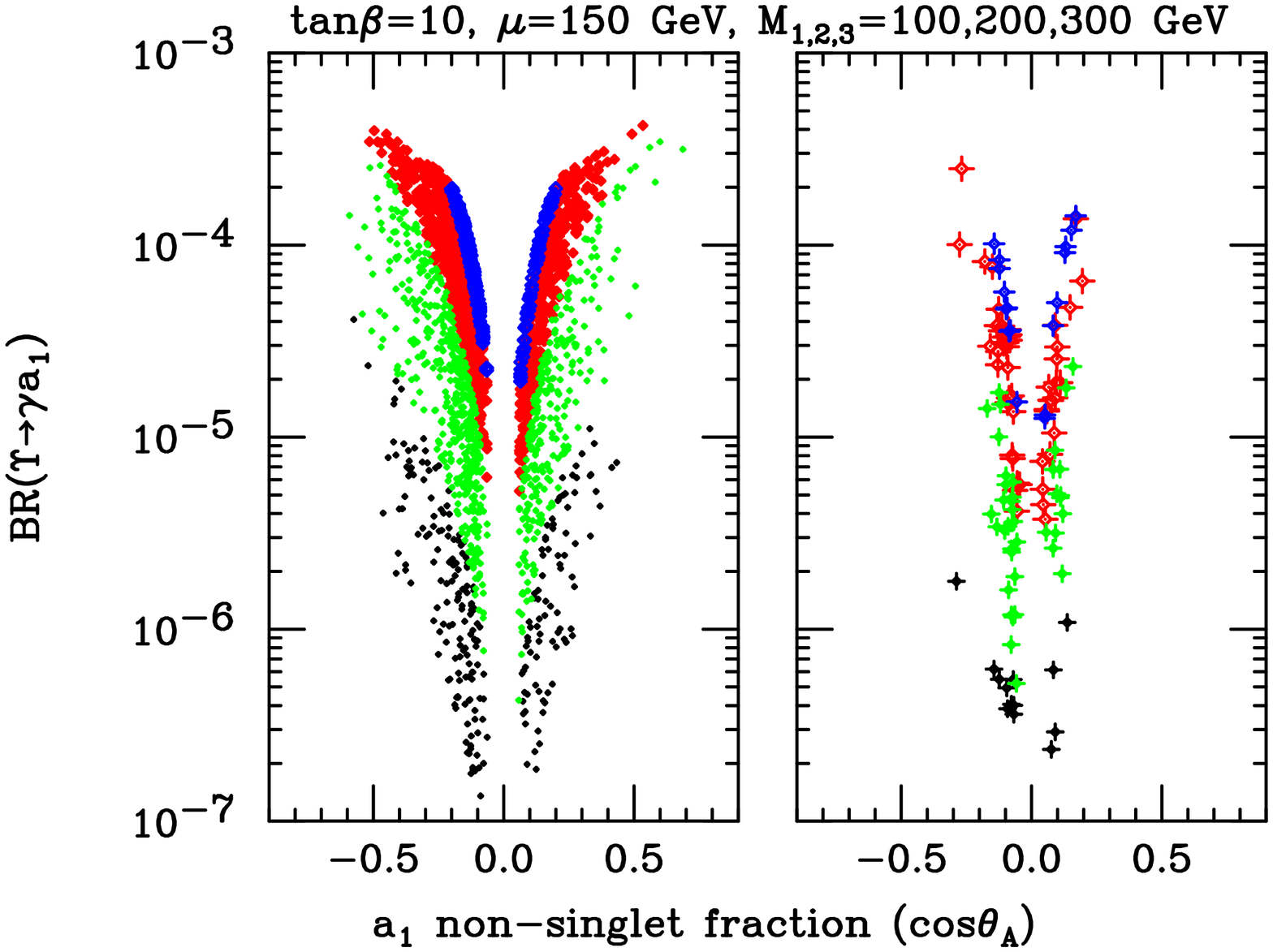}
\caption{$\br(\Upsilon\to\gam\ai)$ for NMSSM scenarios with various
  ranges for $\mai$: dark grey (blue) = $\mai<2\mtau$; medium grey
 (red) = $2\mtau<\mai<7.5\gev$; light grey (green) =
 $7.5\gev<\mai<8.8\gev$; and black = $8.8\gev<\mai<9.2\gev$.
 The plots are for $\tanb=10$ and $M_{1,2,3}(\mz)=100,200,300\gev$.
The left plot comes from the $\alam,\akap$ scan described in the text, holding
$\mueff(\mz)=150\gev$ fixed.  The right plot shows results for $F<15$
scenarios with $\mai<9.2\gev$ found in a general scan over all NMSSM
parameters holding $\tanb$ and $M_{1,2,3}$ fixed as stated.}
\label{upsilon}
\end{figure}

We now turn to $\Upsilon\to \gam\ai$. We have computed the branching
ratio for this decay based on Eqs.~(3.54), (3.58) and (3.60)
of~\cite{hhg} (which gives all appropriate references). Eq.~(3.54)
gives the result based on the non-relativistic quarkonium model;
Eqs.~(3.58) and (3.60) give the procedures for including QCD
corrections and relativistic corrections, respectively.  Both cause
significant suppression with respect to the non-relativistic
quarkonium result.  In addition, there are bound state corrections.
These give a modest enhancement, rising from a small percentage at
small $\mai$ to about $20\%$ at $\mai=9.2\gev$ (see the references in
\cite{hhg}).~\footnote{In contrast, for a scalar Higgs, bound state
  corrections give a very large suppression at higher Higgs masses
  near $\mupsilon$.} For
$\mai\in[\metab-2\Gamma_{\eta_b},\metab+2\Gamma_{\eta_b}]$, where
$\metab\sim\mupsilon-50\mev$ and $\Gamma_{\eta_b}\sim 50\mev$, the
$\ai$ mixes significantly with the $\eta_b$, giving rise to a huge
enhancement of $\brups$.  We have chosen not to plot results for
$\mai>9.2\gev$ since we think that the old theoretical results in this
region require further refinement.  In Fig.~\ref{upsilon}, we present
results for $\brups$ that are consistent with existing experimental
limits~\footnote{We impose the limits of Fig.~3 of
  \cite{Franzini:1987pv}, Fig.~4 of \cite{Albrecht:1985qz}, and
  Fig.~7b of \cite{Albrecht:1985bg}.  The first two limit
  $\br(\Upsilon\to \gam X$), where $X$ is any visible state. The first
  provides the only strong constraint on the $\mai<2\mtau$ region.
  The third gives limits on $\br(\Upsilon\to \gam X)\br(X\to
  \tauptaum)$ that eliminate $2\mtau<\mai<8.8\gev$ points with too
  high $\brups$ (for $\mai>2\mtau$, $\br(\ai\to \tauptaum)\sim 0.9$). 
Since the inclusive photon spectrum from $\Upsilon$ decays falls as
$E_\gam$ increases, the strongest constraints are obtained for 
small $\mai$.}
in two cases: (a) using a scan over $\alam,\akap$ values holding
$\mueff(\mz)=150\gev$ and $M_{1,2,3}(\mz)=100,200,300\gev$ fixed (in
this scan, identical to that described in Ref.~\cite{Dermisek:2006wr},
$\lam$ and $\kap$ are also scanned over and all other SUSY-breaking
parameters are fixed at $300\gev$ -- results are insensitive to this
choice and, therefore, representative of the whole parameter space); (b) for the $F<15$ points found in
the NMSSM parameter scan described earlier.
In both cases, all points plotted pass all NMHDECAY
constraints --- all points have $\mhi\sim 100\gev$, but avoid LEP
constraints by virtue of $\br(\hi\to\ai\ai)>0.7$ and $\mai<2\mb$.  For
both plots, we divide results into four $\mai$ regions: $\mai<2\mtau$,
$2\mtau<\mai<7.5\gev$, $7.5\gev<\mai<8.8\gev$ and
$8.8\gev<\mai<9.2\gev$.  Fig.~\ref{upsilon} makes clear that $\brups$
is mainly controlled by the non-singlet fraction of the $\ai$ and by
$\mai$. The only difference between the (a) and (b) plots is that
$F<15$ restricts the range of $\cta$ to smaller magnitudes (implying
smaller $\brups$) and narrows the $\mai$ bands.  As seen in the
figure, the $\cta\sim -0.05$, $\mai>2\mtau$ scenarios that can have no
$\alam,\akap$ tuning have $\brups\lsim few\times 10^{-5}$.  For
general $\cta$ and $\mai$, values of $\brups$ up to $\sim 10^{-3}$
($5\times 10^{-3}$) are possible for $F<15$ points (in the general
$\alam,\akap$ scan). In Fig.~\ref{upsilon}, points with $\brups\gsim
few\times 10^{-4}$ (depending on $\mai$) are not present, having been
eliminated by 90\% CL limits from existing experiments.  The surviving
points with $\mai<9.2\gev$ can be mostly probed if future running,
upgrades and facilities are designed so that $\brups\sim 10^{-7}$ can
be probed. As stated earlier, predictions at higher $\mai$ are rather
uncertain, but obviously $\brups\to 0$ for $\mai\to\mupsilon$.  To
access higher $\mai$ (but $\mai<2\mb$), $\Upsilon(2S)\to \gam\ai$ and
$\Upsilon(3S)\to \gam\ai$ can be employed; computation of the
branching ratios requires careful attention to $\ai-\eta_b$ mixing,
which can lead to even larger branching ratios than for the $\Upsilon$
if $\mai\sim m_{\eta_b}$.

\begin{figure}
\vspace*{.2in}
\includegraphics[width=2.6in,angle=90]{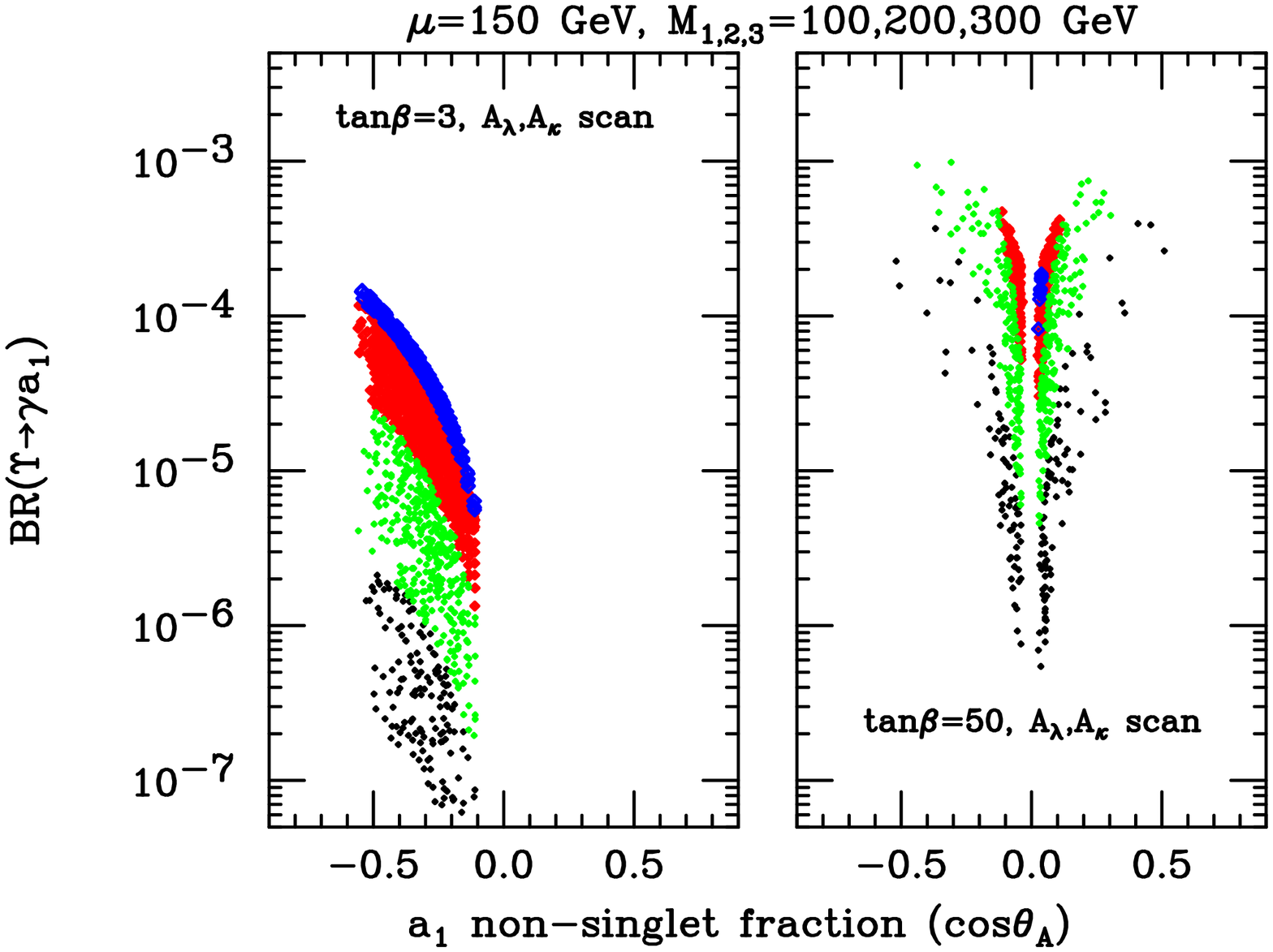}
\caption{We plot $\brups$ as a function of $\cta$ for the
  $\alam,\akap$ scan, taking
  $M_{1,2,3}(\mz)=100,200,300\gev$,
$\mueff(\mz)=150\gev$ with $\tanb=3$ (left) and $\tanb=50$ (right).
The point notation is as in Fig.~\ref{upsilon}.
}
\label{upsilon2}
\end{figure}

Results from the $\alam,\akap$ scan with $\mueff=150\gev$ and
$M_{1,2,3}=100,200,300\gev$ are given in the cases of $\tanb=3$ and
$\tanb=50$ in Fig.~\ref{upsilon2}. Note that
almost all $\tanb=3$ points that pass NMHDECAY and LEP constraints
are consistent with existing limits on $\brups$.
To probe the full set of $\mai<9.2\gev$ points shown, sensitivity to
$\brups\lsim few\times 10^{-8}$ is needed.  Conversely, for $\tanb=50$
a lot of the scan points consistent with NMHDECAY and LEP constraints are already absent because of existing limits
and one need only probe down to $\brups\sim 10^{-6}$ to cover the
$\mai<9.2\gev$ points.

We note that the points with small negative $\cta$ (\eg\ $\cta\sim
-0.5$ for $\tanb=10$) that are most likely to escape $\alam,\akap$
tuning issues are well below the existing limits
from~\cite{Franzini:1987pv,Albrecht:1985qz,Albrecht:1985bg} for all
$\mai$ values for all three $\tanb$ choices.~\footnote{ For a CP-odd
  $a$ that decays into non-interacting states, there are further constraints
  available from Crystal Ball and
  CLEO~\cite{Antreasyan:1990cf}; these only apply to the scenarios
  considered here if $M_1$ is reduced to a very small value (as
  possible without affecting EWSB fine tuning) so that
  $\ai\to\cnone\cnone$ decays are significant. 
  For example, at
  $\tanb=10$, our low fine-tuning scenarios with $M_1$ decreased to
  $3\gev$ can yield $\mcnone\lsim 2\gev$ and $\br(\ai\to \cnone\cnone)\in[0.15,0.35]$. (Generic scenarios with
  substantial $\brups\br(\ai\to \cnone\cnone)$ were considered in~\cite{Gunion:2005rw}.)}  However, none of the
above analyses ~\cite{Franzini:1987pv,Albrecht:1985qz,Albrecht:1985bg}
have been repeated with the larger data sets available from CLEO-III,
BaBar, or Belle. Presumably, much stronger constraints than those we
included can be obtained. Or perhaps a $\gam\ai$ signal will be found.

We expect that the best way to search for the NMSSM light $\ai $ is to use its
exclusive decay modes, as this reduces backgrounds, especially those
important when the photon is soft.  For $\mai > 3.6$ GeV and $\tan
\beta \gsim 1$, the dominant decay mode is $\ai \to \tau^+ \tau^-$.
For example, Ref.~\cite{Sanchis-Lozano:2006gx} has proposed looking
for non-universality in $\Upsilon\to \gam \tauptaum$
vs. $\Upsilon\to\gam \epem,\gam\mupmum$ decays. This
would fit nicely with the low-$F$ scenarios. For $\mai<2\mtau,2m_c $
the decay mode $\ai \to g g$ is generally in the range $20\% - 30\%$,
giving a contribution to $\Upsilon \to \gamma g g$ at the
$10^{-4}$--$10^{-6}$ level; the $s\anti s$ mode is typically larger.

In the $\gamma \tau^+ \tau^-$ final state, the
direct $\gamma \tau^+ \tau^-$ production cross section is 61 pb.
Using signal=background as the criterion, 
this becomes the limiting factor for branching ratios below the $4
\times 10^{-5}$ level when running on the $\Upsilon(1S)$, and below
the $2 \times 10^{-4}$ level when running on the $\Upsilon(3S)$.  To
improve upon the latter, one can select a sample of known $\Upsilon(1S)$
events by looking for dipion transitions from the higher resonances.
The dipion transition gives a strong kinematic constraint on the mass
difference between the two $\Upsilon$'s.  When running on the
$\Upsilon(3S)$, the effective cross section in $\Upsilon(3S) \to \pi^+
\pi^- \Upsilon(1S)$ is 179 pb~\cite{Glenn:1998bd}.~\footnote{This can
  also be done on the $\Upsilon(2S)$ but the pions are softer, implying
  much lower efficiency.  On the $\Upsilon(4S)$ this transition has a
  very small branching ratio $\lsim 10^{-4}$.}  To limit $\brups
 \lsim 10^{-6}$, $5.6 ~{\rm fb}^{-1}/\eps$ would need to
be collected on the $\Upsilon(3S)$, where $\eps$ is the experimental
efficiency for isolating the relevant events.
This analysis can also be done
on the $\Upsilon(4S)$, where the $\Upsilon(3S)$ is produced via ISR.
The effective $\gam_{ISR}\Upsilon(3S)\to \gam_{ISR}\pi^+\pi^-\Upsilon(1S)$ cross
section is 0.78 fb.  To limit
$\brups \lsim 10^{-6}$, $1.3~ {\rm ab}^{-1}/\eps$
would need to be collected. These integrated luminosities
needed to probe $\brups\sim 10^{-6}$ would appear to  
be within reach at existing facilities and would allow discovery
of the $\ai$ for many of the favored NMSSM scenarios.

Are there other modes that would allow direct $\ai$ detection? 
Reference~\cite{Arhrib:2006sx} advocates  $\epem\to \wtil \chi_1^+\wtil \chi_1^- \ai$
with $\ai\to \gam\gam$. This works if the $\ai$ is
very singlet, in which case $\br(\ai\to\gam\gam)$ can be large.
However, see \cite{Dermisek:2006wr} and earlier discussion, a minimum value of $|\cta|$ (\eg\ $|\cta|>0.04$ if
$\tanb=10$) is required in order that $\br(\hi\to\ai\ai)>0.7$
and $\mai<2\mb$.  For the general $\alam,\akap$ scans with 
$\br(\hi\to\ai\ai)>0.7$ and $\mai<2\mb$ imposed, $\br(\ai\to \gam\gam)<4\times
10^{-4}$ with values near $few\times 10^{-5}$ being very common.   It
is conceivable that a super-$B$ factory could detect a signal for
$\Upsilon\to \gam \ai\to \gam\gam\gam$ which would
provide a very interesting check on the consistency of the model.

Flavor changing decays based on $b\to s \ai$ or $s\to d\ai$, in
particular $B\to X_s\ai$, have been examined in
\cite{Hiller:2004ii}. All penguin diagrams containing SM particles
give contributions to the $b\to s \ai$ amplitude that are
suppressed by $\cta/\tanb$ or $\cta/\tan^3\beta$ 
(since up-type quarks couple to the $A_{MSSM}$ with a factor of
$1/\tanb$). Ref.~\cite{Hiller:2004ii} identifies two diagrams involving 
loops containing up-type squarks and charginos that give $b\to s \ai$
amplitudes that are proportional to $\cta\tanb$.
However, the sum of these diagrams vanishes in the super GIM limit
(\eg\ equal up-type squark masses), yielding a tiny $B\to X_s \ai$ transition
rate. Away from this limit, results are
highly model-dependent. In contrast, the predictions for
$\Upsilon\to \gam \ai$ depend essentially only on $\cta$, $\tanb$ and
$\mai$, all of which are fairly constrained for the low-fine-tuning
NMSSM scenarios.

If $\mai<2m_c$, $J/\psi\to \gam \ai$ decay will be possible.  However,
$\br(J/\psi\to \gam\ai)$ is $\sim 10^{-9}$ ($\sim 10^{-7}$) for the smallest
(largest) $|\cta|$ values in the standard $\alam,\akap$ scan for
$\tanb=10$, increasing modestly as $\tanb$ increases.

Before concluding, we note that a light, not-too-singlet $\ai$ could allow
consistency with the observed amount of dark-matter if the $\cnone$ is
largely bino and $2\mcnone\sim \mai$.  This is explored
in~\cite{Gunion:2005rw}.  We found that these scenarios could provide
a consistent description of the dark matter relic density in the case
of a very light $\cnone$.  We report here that this can be coincident
with the $F<15$ scenarios (as well as the small negative $\cta$, 
$\mai>2\mtau$ scenarios that are the most likely to have small
$\alam,\akap$ fine-tuning).  All that is required relative to the $M_1=100\gev$
choice made for our scans is to decrease $M_1$ to bring down $\mcnone$
near $\half \mai$.  $M_1$ is an independent parameter that has
essentially no influence on the value of the fine-tuning measure $F$
so long as $M_1\lsim M_3$.

In summary, aside from discovering the $\ai$ in $\hi\to\ai\ai$ decays,
something that will almost certainly have to await LHC data and,
because of the unusual final state, might not even be seen until the
ILC, it seems that the most promising near-term possibility for testing the
NMSSM scenarios for which EWSB fine-tuning is absent, or more
generally any scenario with large $\br(\hi\to\ai\ai)$ and $\mai<2\mb$,
is to employ the $\Upsilon\to\gam\ai$ decay at either existing $B$
factories or future factories.

\acknowledgments

We are grateful to Miguel Sanchis-Lozano for stressing the importance
of this study as a possible motivation for super-$B$ factories. We
thank M. Peskin and S. Fleming for helpful discussions.  Thanks go to
the Galileo Galilei Institute (JFG) and the Aspen Center for Physics
(JFG and RD) for hospitality and support during the initial stages of
this research.  JFG and BM are supported by DOE grant
DE-FG02-91ER40674 and by the U.C. Davis HEFTI program.  RD is
supported by the U.S. Department of Energy, grant DE-FG02-90ER40542.

\vspace*{-.18in}


\end{document}